\documentclass[prb,aps,twocolumn,epsfig]{revtex4}
\usepackage{epsfig}
\usepackage{color}
\newcommand{\ep}{\epsilon}

\begin{document}
\title{Isolated hybrid normal/superconducting ring in a magnetic flux:
from persistent current to Josephson current}
\author{J\'{e}r\^{o}me Cayssol$^{1}$, Takis Kontos$^{2}$, Gilles Montambaux$^{1}$}
\address{$^{1}$Laboratoire de Physique des Solides, Associ\'e au
CNRS, Universit\'e Paris Sud,  91405 Orsay, France}
\address{$^{2}$CSNSM, CNRS, Universit\'e Paris Sud, 91405 Orsay,
France}
\begin{abstract}
{We investigate the ground state current of an isolated hybrid
normal/superconducting ring (NS ring), threaded by an
Aharonov-Bohm magnetic flux. We calculate the excitation spectrum
of the ring for any values of the lengths of the normal metal and
of the superconductor. We describe the nonlinear flux dependence
of the energy levels above and below the gap edge. Using a
harmonics expansion for the current, we isolate the contribution
due to these nonlinearities and we show that it vanishes for large
normal segment length $d_N$. The remaining contribution is very
easy to evaluate from the linearized low energy spectrum. This
decomposition allows us to recover in a controlled way the
current-flux relationships for SNS junctions and for NS rings. We
also study the crossover from persistent current to Josephson
current in a multichannel NS ring at finite temperature. }
\end{abstract}
%\pacs{PACS Numbers: }
\maketitle

\section{Introduction}
\begin{figure}[ht!]
\begin{center}
 \epsfxsize 6cm
 \epsffile{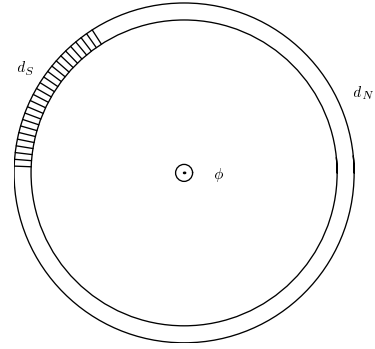}
 \end{center}
\caption{NS ring composed of a normal segment N and a
superconductor segment S, pierced by
  an Aharonov-Bohm flux $\phi$.
}\label{nsringd}
\end{figure}

In spite of the great amount of work devoted to this problem, a
full understanding of the physics of persistent currents in normal
mesoscopic rings is still lacking. On the other hand, the physics
of proximity effect in hybrid normal-superconducting mesoscopic
structures has gained renewed interest recently due to progress in
nanofabrication techniques. What is the interplay between these
two phenomena? In order to address this question, we study a
mesoscopic isolated normal/superconducting loop (NS loop) made of
a normal metal of length $d_N$ and a superconductor of length
$d_S$ as depicted in Fig. \ref{nsringd}. This NS ring is
mesoscopic in the sense that the normal segment is shorter than
the coherence length $L_\phi$. As a consequence of phase
coherence, a non dissipative current flows in the ring when a
magnetic flux $\phi$ is applied. In the absence of superconductor,
i.e. for the normal ring, this is the so-called persistent current
which has the periodicity $\phi_0=h/e$. When the superconducting
segment is longer than its superconducting coherence length
$\xi_o$, this current is analogous to the Josephson current in SNS
junctions, with periodicity $h/2e$.

B\"uttiker and Klapwijk \cite{BK86}(BK) showed that the physical
mechanism responsible for the crossover between these two cases is
the tunneling of Andreev quasiparticules\cite{Andreev64} through
the superconducting segment when its length $d_S$ becomes of order
$\xi_o$. However, BK studied only the low energy spectrum where
energy levels vary linearly with the flux. They constructed the
current-flux relationship $I(\phi)$ in analogy to that of long SNS
junctions. In the present paper, we investigate the full NS loop
spectrum. In particular, we describe carefully nonlinear
variations of the level positions with the flux which appear both
below and above the gap edge. Then, we address the question of a
proper calculation of the current $I(\phi)$ which includes these
new spectral features. The standard difficulty is to compute the
sum of many single level currents $-\partial \ep/\partial \phi$
which are of the same order of magnitude and alternate in sign. In
our approach, each harmonic of the total current $I(\phi)$ is
expressed as the sum of a term proportional to a single level
current at zero energy $-\partial \ep/\partial \phi(\ep=0)$ plus
an integral of the "curvature" $-\partial^{2} \ep/\partial^{2}
\phi$ over the whole spectrum. The latter term comes specifically
from the nonlinearities in the spectrum and we show numerically
that it is significant only for short normal segment, typically
when $d_N\lesssim 2 \xi_o$. For longer normal segment $d_N \gtrsim
2 \xi_o$, the current $I(\phi)$ can be obtained simply from the
low energy linearized spectrum. As a byproduct, one recovers the
result for the long SNS junction when $d_{S} \rightarrow \infty$
and for the normal ring when $d_S \rightarrow 0$. This derivation
of $I(\phi)$ for long SNS junctions  is surprisingly simple
compared to the original derivation of Bardeen-Johnson, Ishii and
Svidsinski {\it et al.}
\cite{Bardeen72}$^,$\cite{Ishii70}$^,$\cite{Svid72}. These authors
found that $I(\phi)$ is triangular, but disagreed with the value
of the critical current. The simplicity of our approach enables us
to show that Bardeen and Johnson found the correct critical
current for long SNS junctions. Beside the simplicity of the
derivation, our decomposition gives us the possibility to control
the approximation since the 'curvature' term is a correction to
these well-known results. For $d_N=0$, we can evaluate both terms
and by summation, we reconstruct the short junction result: the
current is $2 \pi\Delta/\phi_o \sin(\Delta\chi/2)$ per channel
\cite{KO}$^,$\cite{Beenakker1}. As a new result, we study the
crossover from the persistent current to the Josephson current at
finite temperature for single-channel and for multichannel NS
loops.

The paper is organized as follows: in section \ref{sexcit}, we
recall the expression of the thermodynamic potential in terms of
the excitation spectrum and we derive the excitation spectrum of a
single-channel NS loop for arbitrary $d_N$ and $d_S$. In section
\ref{sdecomp}, we derive our new decomposition of the current and
evaluate the contribution of the nonlinear flux dependent energy
levels. In section \ref{slongdn}, we apply this approach to
demonstrate the BK result valid in the case of a long normal
segment and we consider the evolution of the full spectrum when
$d_S$ varies. Section \ref{sshortdn} describes the crossover from
long SNS junctions to short junctions as $d_N$ is decreased.
 The contributions to the ground state current
from levels above and below the gap $\Delta$ is discussed in
section \ref{sinout}. Finite temperature effects are incorporated
in section \ref{stempe} to study the transition from a diamagnetic
to a paramagnetic behavior at small flux when $d_S$ is lowered or
when $T$ is increased. Finally, in section \ref{mulcanal} shows
how to sum over transverse channels to obtain the multichannel
case from our study of the single-channel case. We conclude in
section \ref{conclu}.

\section{Excitation spectrum and supercurrent of the NS loop}\label{sexcit}

\subsection{Relation between supercurrent and excitation spectrum}
\label{ssexcit1}
\begin{figure}[ht!]
\centerline{ \epsfxsize 4cm \epsffile{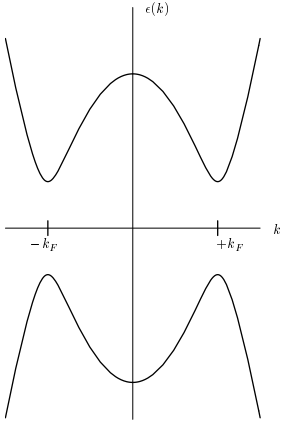}}
 \caption{Excitation spectrum (top curve) versus semiconductor model spectrum (top+bottom curves)
 of a large purely superconducting ring.}\label{excitation}
\label{modelsemi}
 \end{figure}

We consider a NS loop of perimeter $L$ made of a
superconducting segment of length  $d_S$ and a normal segment of
length $d_N$. The non-dissipative current flowing in this system
is obtained by differentiation of the thermodynamic potential or
Gibbs energy $\Omega(T,\mu,\phi)$ with respect to the magnetic
flux:
\begin{equation}
I(\phi) = -\left({\partial \Omega \over \partial
\phi}\right)_{\mu,T}\label{IAphi}
\end{equation}
For a system with inhomogeneous superconductivity, Bardeen $et$
$al.$ and Beenakker $et$ $al.$ have shown that it is possible to
express the thermodynamic potential in terms of the excitation
spectrum \cite{Bardeen}$^{,}$\cite{Beenakker2}. This excitation
spectrum is found by solving the Bogoliubov-De Gennes equations
\cite{Gennes}:
\begin{equation}
\left(
\begin{array}{cc}
H_{o}     &  \Delta(x) \\ \Delta^{*}(x) & -H_{o}^{*}\\
\end{array}
\right)
\left(
\begin{array}{c}
 u (x)\\ v(x)
\end{array}
\right) =\ep \left(
\begin{array}{c}
 u (x)\\ v(x)
\end{array}
\right)\label{Bdgeq}
\end{equation}
These equations apply when the normal segment is shorter than the
phase coherence length $L_{\phi}$. For such a mesoscopic NS ring,
excitations are coherent around the whole loop and can be
described by electron-like and hole-like wavefunctions denoted
respectively by $u(x)$ and $v(x)$. $H_{o} =(-i \hbar d/dx -q
A(x))^2/2m+V(x)-\mu$ where $\mu$ is the chemical potential and $m$
is the effective mass of electrons and holes common for both
superconducting and normal parts. $A(x)$ is the vector potential
due to the Aharonov-Bohm flux, $V(x)$ is the disorder potential
and $x$ is the coordinate along the loop. In this paper, we
consider the clean system $V(x)=0$. Following BK, we choose a
"square well" model for the superconducting gap: $\Delta(x)$ is
zero in the normal region and uniform $\Delta(x)=\Delta$ in the
superconductor. The thermodynamic potential can be written in
terms of the excitation energies and of the superconducting gap:

\begin{eqnarray}
\Omega(T,\mu,\phi)&=& - 2T \int_{0}^{\infty} d\epsilon \ln(2 \cosh
\frac{\ep}{2T}) \rho_{exc}(\ep,\phi) \nonumber \\ &+& \int dx
|\Delta(x)|^2 / g + Tr H_o \label{Potexcit}
\end{eqnarray}
where $g$ is the pairing interaction present in the
superconducting segment. In front of the first integral, the
factor 2 accounts for spin degeneracy and we choose units with
$k_{B}=1$. In this formula, $\rho_{exc}(\ep,\phi)$ is the density
of excited states per spin direction. The last two terms in Eq.
(\ref{Potexcit}) are independent of the flux. The first term can
be interpreted as the Gibbs energy for the semiconductor model. In
the semiconductor model, states with positive energies lie at the
quasiparticle energies of the initial problem and states below the
Fermi level lie at the opposite of the former quasiparticles
energies, as depicted in Fig. \ref{modelsemi}. By construction,
the spectrum of this semiconductor model is fully symmetric with
respect to its Fermi level and each state can only be singly
occupied. For this semiconductor, the flux dependent part of the
thermodynamical potential is:
\begin{equation}
\Omega(T,\mu,\phi)= - T \int_{-\infty}^{\infty} d\ep \ln (1+\exp
-\frac{\ep}{T}) \rho(\ep,\phi) \label{Potsemi}
\end{equation}
where $\rho(\ep,\phi)=\rho_{exc}(|\ep|,\phi)$ is the semiconductor
density of states obtained by symmetrisation of the original
density of excited states as represented in Fig. \ref{modelsemi}.

\subsection{Excitation spectrum for arbitrary NS loop}\label{ssexcit2}

\subsubsection{Derivation of the spectrum}
In this section, we calculate the flux-dependent excitation
spectrum $\ep(\phi)$ of the NS loop. BK have established this
excitation spectrum below the gap $\Delta$ and for a fixed number
$N$ of electrons per spin direction\cite{BK86}. Here, we calculate
the spectrum below and above $\Delta$, considering as well the
cases of a fixed number of particles or a fixed chemical potential
$\mu=k_{F}^2/2m$. The case of a fixed number of electrons per spin
direction corresponds to $k_{F}L=\pi N$ \cite{numberel}. The
excitation energies are found by solving the Bogoliubov-De Gennes
equation (\ref{Bdgeq}) for the NS loop. The eigenvalue equation is
obtained by matching the two-component wavefunctions and their
derivatives at the NS interfaces, see Appendix \ref{app1}. In the
quasiclassical approximation $\ep \ll E_F$, the resulting equation
is:
\begin{equation}
\cos k_{F}L=r_{\ep} \cos \left(\frac{\ep d_N}{\hbar v_F}\mp 2 \pi
\varphi + \Theta_{\ep} \right)\label{Bktrig}
\end{equation}
where $\varphi=\phi/\phi_o$ is the reduced flux. The minus sign
corresponds to excitations around $+k_F$ and the plus sign to
excitations around $-k_F$. $\Theta_\ep$ is a phase shift due to
the presence of the superconductor that adds to the phase shift
associated to the motion in the normal segment. The function
$r_\ep$ defines an energy dependent renormalization of the Fermi
wavevector $\cos k_{F}L\rightarrow\cos k_{F}L/r_\ep$. The
functions $r_\ep$ and $\Theta_\ep$ have different expressions
below and above the gap.
Inside the gap $\ep<\Delta$, they are
given by, see appendix \ref{app1}:
\begin{equation}
r_{\ep} e^{i \Theta_{\ep}}=\frac{\sinh(2 i
\eta_{\ep}-\lambda_{\ep} d_{S})}{\sinh 2 i \eta_{\ep}}
\label{compl1}
\end{equation}
where $e^{2i\eta_{\ep}}=( \ep + i\sqrt{\Delta^2-\ep^2})/\Delta$
and $\lambda_{\ep}=\sqrt{\Delta^2-\ep^2}/\hbar v_F$. At zero
energy, $\lambda=\lambda_{\ep=0}$ is the inverse of the
superconducting coherence length $\xi_o=\hbar v_F / \Delta$ which
is the characteristic length scale for this system
\cite{cohlength}. The modulus of the complex number (\ref{compl1})
is:
\begin{equation}
r_{\ep}=\mid \cosh \lambda_{\ep} d_S+ i \cot 2 \eta_{\ep} \sinh
\lambda_{\ep} d_S \mid
\label{rep1}
\end{equation}
and the phase $\Theta_{\ep}$ satisfies:
\begin{equation}
\tan \Theta_{\ep}=\cot 2 \eta_{\ep} \tanh \lambda_{\ep} d_S
\label{tanthet}
\end{equation}
Above the gap, Eq. (\ref{compl1}) becomes, see appendix
\ref{app1}:
\begin{eqnarray}
r_{\ep} e^{i \Theta_{\ep}}&=&\frac{\sinh(i \delta k_{\ep} d_{S}+ 2
\delta_{\ep})}{\sinh 2 \delta_{\ep}}\\ &=& \cos \delta k_{\ep}
d_{S} + i \coth 2 \delta \sin \delta k_{\ep} d_{S}
\label{compl2}
\end{eqnarray}
so that $\Theta_\ep$ and $r_\ep$ satisfy:
\begin{equation}
r_{\ep}=\mid \cos \delta k_{\ep} d_{S} + i \coth 2 \delta_{\ep}
\sin \delta k_{\ep} d_{S} \mid
\label{rep2}
\end{equation}
and:
\begin{equation}
\tan \Theta_{\ep}= \coth 2 \delta_{\ep} \tan \delta k_{\ep} d_{S}
\label{t2a}
\end{equation}
where $e^{2\delta_{\ep}}=(\ep + \sqrt{\ep^2-\Delta^2})/\Delta$ and
$\delta k_{\ep}=\sqrt{\ep^2-\Delta^2}/\hbar v_F$. We choose
$\Theta_\ep$ to have the same integer part as $\delta k_{\ep}
d_S$. This implies:
\begin{equation}
\Theta_{\ep}=\arctan \left(\frac{\tan \delta k _{\ep} d_S}{\tanh 2
\delta_{\ep}} \right)+ \pi Int \left(\frac{\delta k _{\ep}
d_S}{\pi}+\frac{1}{2} \right) \label{Thetaup}
\end{equation}
The spectrum above the gap $\ep > \Delta$ was not considered in
the work of BK. Excitation energies are thus solutions of the
quantization equation (\ref{Bktrig}) valid whether the energy is
above or below $\Delta$. The complexity is hidden in the energy
dependence of the functions $r_\ep$ and $\Theta_\ep$ given
respectively by Eqs. (\ref{rep1},\ref{rep2}) and by Eqs.
(\ref{tanthet},\ref{t2a},\ref{Thetaup}). These functions are
plotted in the Figs. (\ref{repsi},\ref{figthet1},\ref{figth2}) of
Appendix \ref{app1}. The correspondence between the solutions
above and below the gap is given by the mapping:
\begin{eqnarray}
-i \eta_{\ep} &\longrightarrow& \delta_{\ep} \nonumber \\
\lambda_{\ep} &\longrightarrow& i \delta k_{\ep}
\end{eqnarray}
The excitation energies are the positive solutions of Eq.
(\ref{Bktrig}). They are quantized according to $\ep^j(n \pm
\varphi)$, the two fonctions $\ep^j(y)$ being:
\begin{equation}
\ep^{j}(y) = {{h v_F} \over d_N }\left[ y - {\Theta_\ep \over 2
\pi}+{ j \over 2 \pi} \arccos \left(\frac{\cos
k_{F}L}{r_\ep}\right) \right] \label{epy1}
\end{equation}
where $j=\pm 1$.
\begin{figure}[ht!]
\begin{center}
\epsfxsize 9cm \epsffile{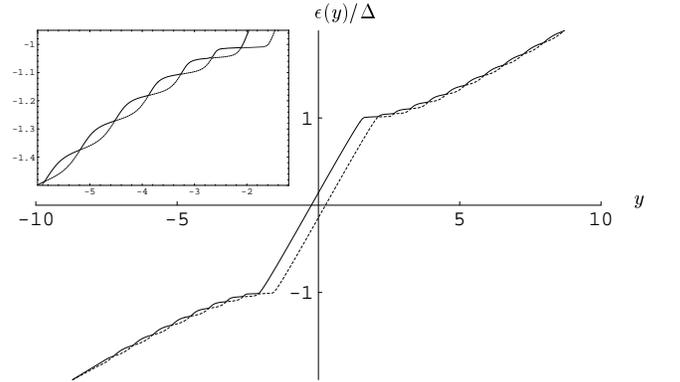}
\end{center}
\caption{$\epsilon^{j}(y)$ for $j=+1$ (solid line) and $j=-1$
(dashed line) in a NS loop with $d_{S}=20 \xi_o$, $d_{N}=10
\xi_{o}$, containing an even number of electrons per spin
direction.} \label{yepgen}
\end{figure}
All the information about the spectrum and its flux dependence is
contained in the functions $\ep^{j}(y)$. We use these functions to
calculate the current in the following sections. As an example, we
have plotted  in Fig. \ref{yepgen} the two functions $\ep^{j}(y)$
for a NS loop with $d_{S}=20\xi_o$ and $d_{N}=10\xi_o$ containing
an even number of particles per spin direction.

\subsubsection{Linear and nonlinear regions of the NS loop spectrum}
The flux dependent spectrum is obtained by folding the curves
$\ep^{j}(y)$ in the interval $[-1/2,1/2]$. It is shown on Fig.
\ref{spectregen} for the example $d_{S}=20\xi_o$ and
$d_{N}=10\xi_o$. At zero flux, the first energy levels
$\ep^{j}(n,\varphi=0)$ correspond to
$(n,j)=(0,1),(1,-1),(1,1),(2,-1),$ etc. Now, we examine the
general structure of the excitation spectrum for arbitrary values
of $d_{N}$ or $d_{S}$. We distinguish three regions:

\begin{figure}[ht!]
\begin{center}
\epsfxsize 8cm \epsffile{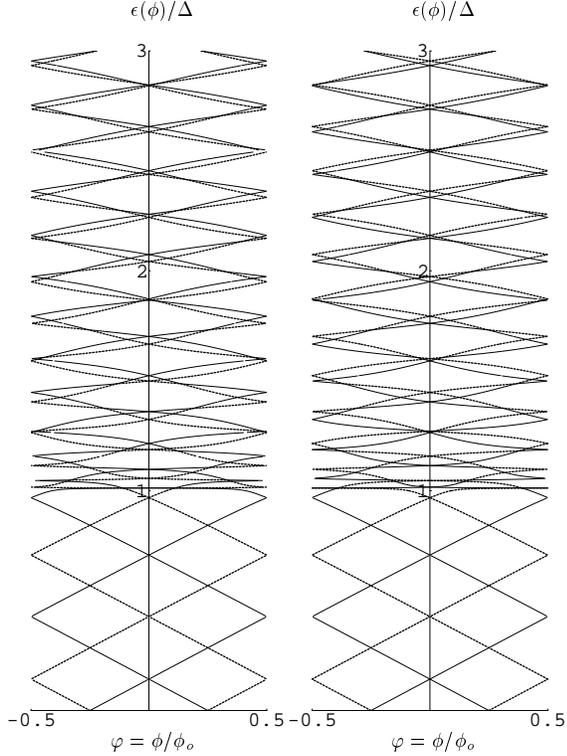}
\end{center}
\caption{NS loop spectrum $\ep^{j}_{\pm}(n,\varphi)$ for $j=+1$
(thin line) and $j=-1$ (thick line) $d_{S}=20 \xi_o$, $d_{N}=10
\xi_{o}$ for an even (left) and for an odd (right) number of
electrons $N$ per spin direction. For large energies, the two
branches $j=\pm 1$ tend to coincide. The high energy levels
exhibit a parity effect and are close to those of the normal ring.
The Andreev levels are insensitive to the parity of $N$.
}\label{spectregen}
\end{figure}

\begin{figure}[ht!]
\begin{center}
\epsfxsize 8cm \epsffile{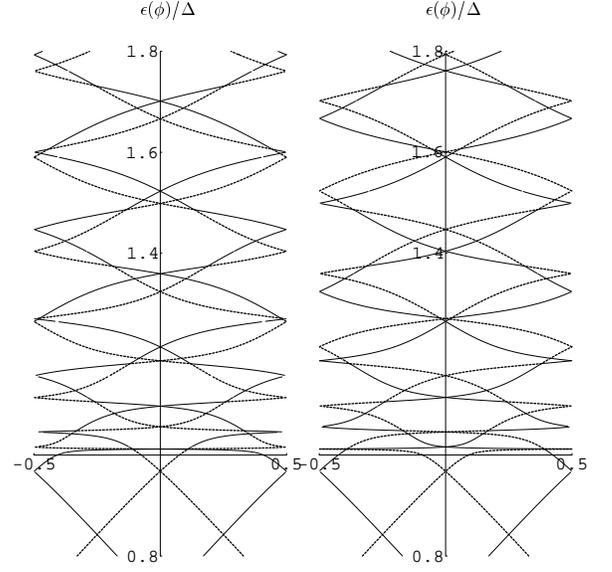}
\end{center}
\caption{NS loop spectrum $\ep^{j}_{\pm}(n,\varphi)$ in the vicinity
of $\Delta$ for $j=+1$ (thin line) and $j=-1$ (thick line)
$d_{S}=20 \xi_o$, $d_{N}=10 \xi_o$ and $N$ even (left) or odd
(right).}\label{spectrezoom}
\end{figure}

i) The low energy spectrum $\ep < \Delta$: in this limit,
$\Theta_{\ep} \simeq \ep/\Delta . \tanh \lambda d_{S}$ and
$r_{\ep} \simeq \cosh \lambda d_{S}$. The resulting form of the
spectrum equation (\ref{Bktrig}) is:
\begin{equation}
\frac{\cos k_{F}L}{\cosh\lambda d_{S}} \simeq \cos \left(\frac{\ep
d^{*}}{\hbar v_F}\mp 2 \pi \varphi \right)\label{Bktriglow}
\end{equation}
and the energy levels vary linearly with the magnetic flux:
\begin{equation}
\ep^{j}_{\pm}(n,\varphi) = \frac{h v_F}{d^{*}} \left[n \pm \varphi +
\frac{j}{2\pi} \arccos \left({\cos k_{F}L \over \cosh\lambda d_S}
\right)\right] \label{bklinspec}
\end{equation}
where $d^{*}=d_N + \xi_{o} \tanh \lambda d_{S}$ is the effective
size of the normal segment. The low energy flux dependent spectrum
is made of linear sections with slopes $\pm e v_{F}/d^*$. There
are two sets of levels corresponding to $j=+1$ and $j=-1$. Each
set is made of equally spaced levels with a common average spacing
$h v_{F} / 2 d^*$. The energy shift between the two sets is $h
v_{F} / \pi d^{*}. \arccos(\cos k_{F}L/\cosh\lambda d_{S})$. For
$d_{S}\gg\xi_o$, this shift is $h v_{F}/ 2 d^{*}$. Taking into
account the spin degeneracy, there are $4 d^{*}/\pi \xi_o$
quasiparticle states inside the gap.

ii) The high energy spectrum $\ep \gg \Delta$: then $\Theta_{\ep}
\simeq \delta k_{\ep} d_S \simeq \ep d_{S}/\hbar v_{F}$ and
$r_{\ep} \simeq 1$. Eq. (\ref{Bktrig}) takes the asymptotic form:
\begin{equation}
\cos k_{F}L \simeq \cos \left(\frac{\ep L}{\hbar v_F}\mp 2 \pi
\varphi \right)\label{Bktrighigh}
\end{equation}
and the spectrum is linear in flux:
\begin{equation}
\ep^{j}_{\pm}(n,\varphi)=\frac{h v_F}{L}  \left[ n \pm \varphi +
\frac{j}{2 \pi} \arccos(\cos k_{F}L ) \right]
\label{Normalspectrum}
\end{equation}
The high energy spectrum has the same linear structure as the low
energy spectrum but with a smaller slope $\pm e v_{F}/L$ which
corresponds to excitations extended around the whole loop. Indeed,
Eq. (\ref{Normalspectrum}) is the linearized excitation spectrum
of a purely normal ring of perimeter $L=d_{S}+d_{L}$. For a given
parity, $\cos k_{F}L=\pm 1$, the two sets of levels $j=+1$ and
$j=-1$ are in coincidence, as depicted in Fig. \ref{spectregen}.

iii) Between the regions i) and ii), the curvature $\partial^{2}
\ep/\partial^{2}\phi$ is finite and alternates in sign. It is
impossible to linearize the spectrum in this region plotted in
Fig. \ref{spectrezoom}. One has to be very careful in evaluating
the current carried by these levels. This will be done in section
\ref{sdecomp}.

\subsection{From the normal ring to the SNS junction}\label{ssexcit3}

In the limit $d_{S}/\xi_{o} = 0$, we have $\Theta_{\ep}=0$ and
$r_{\ep}=1$ and we recover the quantization condition for a normal
ring of length $L$:
\begin{equation}
\cos k_{F}L=\cos \left(\frac{\ep L}{\hbar v_F}\mp 2 \pi
\varphi \right)\label{Bktrighigh2}
\end{equation}
The corresponding linearized spectrum is:
\begin{equation}
\ep^{j}_{\pm}(n,\varphi)=\frac{h v_F}{L}  \left[ n \pm \varphi +
\frac{j}{2 \pi} \arccos(\cos k_{F}L ) \right]
\label{Normalspectrum2}
\end{equation}
characteristic of a purely normal ring \cite{Bloch}. In Eq.
(\ref{Normalspectrum2}), $\pm \varphi$ stands for excitations with
momentum around $\pm k_F$. The index $j=1$ corresponds to
hole-like excitations with $\mid k\mid<k_F$ and $j=-1$ to
electron-like excitations $\mid k\mid>k_F$.

In the opposite limit, for large $d_{S}/\xi_o$, $1/r_\ep$ is
vanishingly small in the gap region and $\Theta_\ep$ is simply
related to the Andreev energy dependent phase shift at a NS
interface by $\Theta_{\ep}=\pi/2-\arccos(\ep/\Delta)$.
Consequently, the Andreev level spectrum given by Eq. (\ref{epy1})
becomes:
\begin{equation}
\ep_{\pm}^{j}(n,\varphi)=\frac{h v_F}{d_N} \left[n \pm \varphi +
\frac{1}{2 \pi } \arccos \frac{\ep}{\Delta} + \frac{j-1}{4}
\right]\label{kuliknj}
\end{equation}
We can express this latter spectrum with only one quantum number
$m=2n+(j-1)/2$, the first excited states at zero flux
corresponding to $m=0,1,2,$etc:
\begin{equation}
\ep_{\pm}(m,\varphi)=\frac{h v_F}{2 d_N} \left[m \pm 2\varphi +
\frac{1}{\pi } \arccos \frac{\ep}{\Delta}
\right]\label{Kulikspectrum}
\end{equation}
We recognize in Eq. (\ref{Kulikspectrum}) the spectrum discovered
by Kulik for bound states in a SNS junction \cite{Kulik69}, the
difference between the phases $\chi_1$ and $\chi_2$ of the
superconducting order parameters of the leads being
$\chi_{1}-\chi_{2}=4 \pi \varphi=4\pi\phi /\phi_o$. This spectrum
can be understood in terms of quantization along closed orbits in
which one electron propagates in one direction along the normal
segment, is reflected as a hole at the superconductor interface
with a phase shift $\chi_{1S} - \arccos(\ep/\Delta)$, then the
hole goes back along the normal segment and is finally reflected
as an electron with an additional phase shift $-\chi_{2S} -
\arccos(\ep/\Delta)$. This explains why the level spacing at
$\chi_{1S}=\chi_{2S}$ is $ h
 v_F / 2 d_N$ corresponding to a box of size $2 d_N$. This
scheme applies because the electron cannot tunnel through the
superconductor in the regime $d_{S}\gg \xi_o$.

\section{Harmonic expansion of the supercurrent}\label{sdecomp}

In the previous section, we have shown that the NS spectrum can be
linearized for low and for high energies compared to $\Delta$.
Close to the gap edge, we have identified a complicated nonlinear
variation of the levels with the flux. Instead of the usual
decomposition between current carried by the Andreev levels and by
the levels above the gap\cite{Bagwell}$^{,}$\cite{Samuel}, we
demonstrate that one can extract a contribution to the current
harmonics specifically due to the nonlinearities, namely a term
proportional to $\partial^{2}\ep/\partial^{2}\phi$. Here, the
nonlinearities occur because a quasi-particule experiences an
energy dependent phase shift when it goes through the NS boundary.
Our approach is quite general and can be applied for other systems
where a nonlinear behavior occurs. As an example, in appendix
\ref{appkfl}, we use this formalism to calculate the correction to
persistent current of the normal ring with a quadratic dispersion
relation.

\subsection{Derivation of the main result}

We start within the framework of the semiconductor model and we
use the functions $\ep^j(\varphi)$ introduced in the previous
paragraph. In order to simplify the notations, we first write the
current for one value of $j$ and omit the index $j$ for
convenience. Hence, we consider the spectrum $\ep(n\pm\varphi)$
and $n=...,-1,0,1,..$ in the semiconductor model representation.
We can express the Gibbs energy (\ref{Potsemi}) in terms of the
double integral of the density of states $N(\ep,\phi)$ defined by:
\begin{equation}
N(\ep,\phi)=\int_{-\infty}^{\ep}d\ep'\int_{-\infty}^{\ep'} d\ep''
\rho(\ep'',\phi)
\end{equation}
in the following manner:
\begin{equation}
\Omega(T,\mu,\phi)= \int_{-\infty}^{\infty} d\ep N(\ep,\phi)
\left( {\partial f } \over {\partial \ep} \right)
\end{equation}
leading to the current given by Eq. (\ref{IAphi}):
\begin{equation}
I(\phi) =\int_{-\infty}^{\infty} {{d\ep} \over{4 T \cosh^{2} \ep /
2T }}\left({\partial N(\ep,\phi) \over
\partial \phi}\right)_{\mu} \label{In2phi}
\end{equation}
The quantities $\rho(\ep,\phi)$ and $N(\ep,\phi)$ are even
functions of the magnetic flux. Omitting the flux independent
part, we write the Fourier decomposition of $N(\ep,\phi)$ as:
\begin{equation}
N(\ep,\phi)= \sum_{{m}=1}^{\infty} N_{m}(\ep) \cos {2 \pi {m}
 \varphi} \label{NDosFourier}
\end{equation}
The density of states of the semiconductor model is given by:
\begin{equation}
\rho(\ep,\phi)=\sum_{n=-\infty,\sigma=\pm 1}^{n=\infty}
\delta(\ep-\ep(n+\sigma \varphi))\label{Dosdelta}
\end{equation}
Using the Poisson summation formula, one gets the Fourier
harmonics of the density of states :
\begin{equation}
\rho_m(\ep)=4 \int_{-\infty}^\infty dy \cos 2 \pi m y \delta(\ep -
\ep(y))
\end{equation}
in terms of $\ep(y)$, which is given by  Eq. (\ref{epy1}) in the
NS loop problem. After a double integration, one obtains the
coefficients $N_m(\ep)$ :
\begin{equation}
N_{m}(\ep)=4 \int_{-\infty}^{y(\ep)} dy' \frac{\sin 2 \pi m y'}{2
\pi m} \frac{d\ep(y')}{dy'} \label{N1m}
\end{equation}
Finally, the current is given by:
\begin{equation}
I(\phi)= \sum_{{m}=1}^{\infty} I_{m} \sin {2 \pi {m} \varphi}
\label{IFourier}
\end{equation}
with:
\begin{equation}
I_{m}(T) = - \frac{2 \pi m}{\phi_o} \int_{-\infty}^{\infty}
{{d\ep} \over{4 T \cosh^{2} \ep/2T }} N_{m}(\ep) \label{ImNmT}
\end{equation}
At $T=0$, the current harmonics are given by $I_{m}(T=0) = - 2 \pi m
N_{m}(\ep=0)/{\phi_o}$. Integrating by parts Eq. (\ref{N1m}), one
gets :

\begin{eqnarray}
I_{m}(T&=&0)= {2 \over {\pi m}} \frac{1}{\phi_o} \left[ {d\ep
\over dy} (y_{o}) \cos 2 \pi m y_{o}\right. \nonumber \\&-& \left.
\int_{-\infty}^{ y_{o}} dy \frac{d^2\ep}{d^2y} \cos 2 \pi m y
\right]  \label{Im}
\end{eqnarray}
with $y_{o}=y(\ep=0)$. We have assumed that the slope $d\ep/dy$ is
vanishing at the bottom $y=-\infty$ of the semiconductor valence
band. Eq. (\ref{Im}) is the general expression of the current for
a spectrum $\ep(n\pm\varphi)$. For the NS loop case, we have to
sum contributions from the two branches of levels $j=\pm 1$. To
understand Eq. (\ref{Im}), we recall that $y$ plays the role of
the reduced flux $\varphi=\phi/\phi_o$. The first term is related
to the slope of $\ep(y)$ at zero energy, i.e. to the current
$-\partial\ep/\partial\phi$ carried by the zero energy Andreev
level. It leads to a triangular $I(\phi)$ current-flux
relationship. The second term is a sum over the whole spectrum of
an integrand proportional to $d\ep/dy$, i.e.
$\partial^2\ep/\partial^{2}\phi$. Only the region around the gap
edge with nonlinearities gives a non zero contribution, regardless
whether these nonlinearities are located below or above $\Delta$.
For this reason, our representation of the current is different
from the usual decomposition of the Josephson current for SNS
junctions as a contribution from the discrete spectrum below the
gap plus a contribution from the continuum spectrum above the gap
\cite{Bagwell}.

\subsection{Numerical evaluation.}\label{snum}
In this section, we show that the relative weight of the two terms
in Eq. (\ref{Im}) is related to the value of $d_N$. We evaluate
numerically the integral term in Eq. (\ref{Im}) for loops with
finite lengths $0\leq d_{N}\leq10\xi_o$ and $0\leq
d_{S}\leq10\xi_o$.
\begin{figure}[ht!]
\begin{center}
 \epsfxsize 8.5cm \epsffile{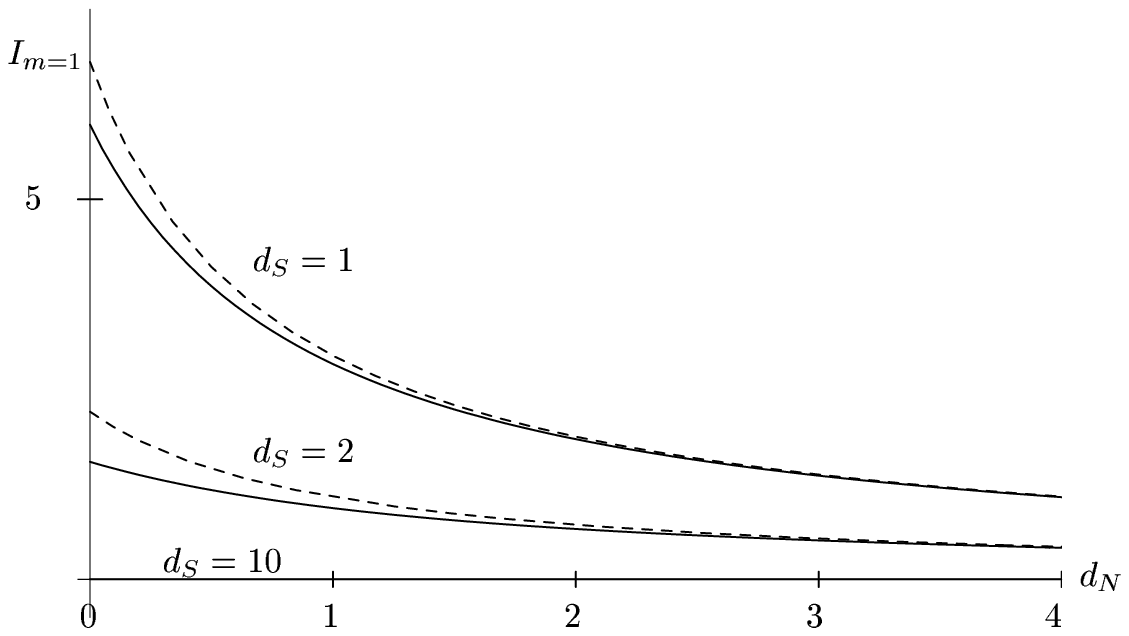}
\end{center}
\begin{center}
 \epsfxsize 8.5cm
\epsffile{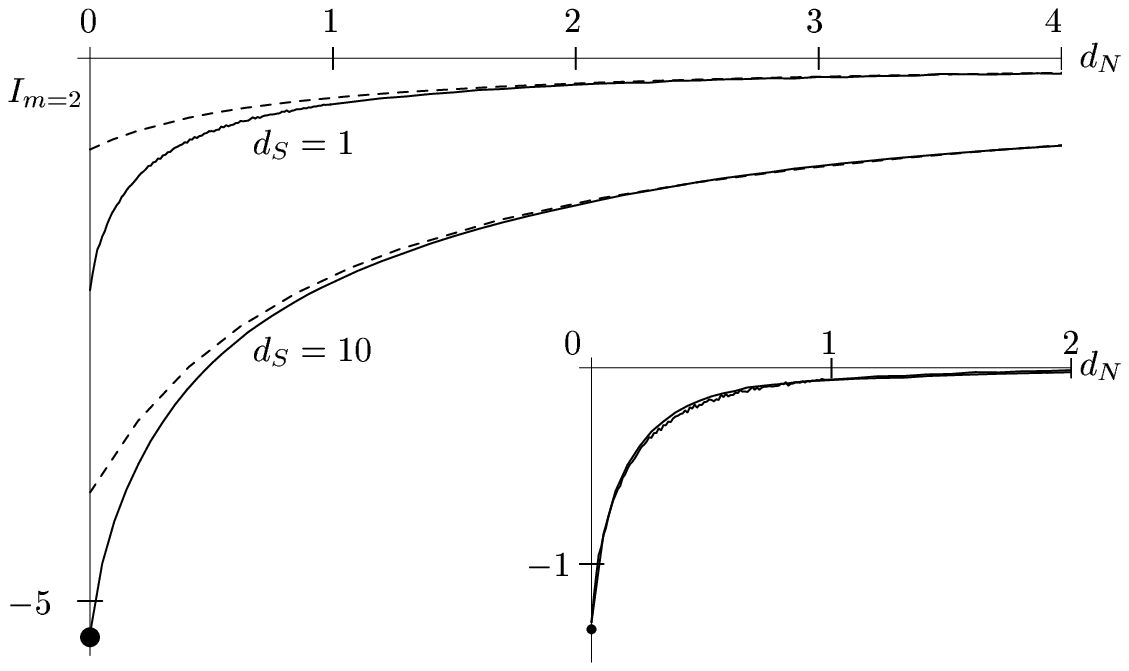}
\end{center}
\caption{First harmonic $I_{m=1}$ (top curve) and second harmonic
$I_{m=2}$ (bottom curve) in units of $\Delta/\phi_o$ as a function
of $d_{N}$, for different values of $d_S$. Dashed lines represent
the BJ approximation. Inset shows the difference between the
second harmonic and its BJ evaluation for $d_S=1$ and $d_S=10$.
The dot represents the expected value for $d_{N}=0$ and $d_{S} \gg
\xi_o $. } \label{harmo}
\end{figure}
In Fig. (\ref{harmo}), we have plotted the first $I_{m=1}$ and
second $I_{m=2}$ harmonics of $I(\phi)$ as a function of $d_N$,
for different values of $d_{S}$. The dashed lines represent the
first term in Eq. (\ref{Im}) which corresponds to the
Bardeen-Johnson(BJ) approximation used by BK. One sees that the
second term is already negligible for $d_{N} \gtrsim 2$. For small
values of $d_N$, both terms in Eq. (\ref{Im}) are of the same
order of magnitude. The difference between the exact result and
the BJ approximation, which is shown in the inset of Fig.
\ref{harmo} for $I_{m=2}$, decreases monotonously with increasing
$d_N$. We note that this correction to the BJ approximation is
roughly independent of $d_S$ for $m=2$. As it is expected for odd
harmonics, $I_{m=1}$ decreases as $d_S$ increases, see Fig.
\ref{harmo}. For $d_{S}=10 \xi_o$, our numerical evaluation of the
second harmonic is in good agreement with the analytical value
$I_{m=2}=-16 \Delta/3\phi_o$ expected for $d_{N}=0$ and
$d_{S}=\infty$ from Eq. (\ref{ipshort}).

\subsection{Conclusion}
We have identified a term specifically due to nonlinerities in the
spectrum and we have shown numerically that it is small if $d_N
\lesssim 2 \xi_o$. In the following sections, we use Eq.
(\ref{Im}) to recover analytical expressions in the extreme cases:
$d_{N}\gg\xi_o$ for any $d_S$ (section \ref{slongdn}) and
$d_{S}\gg\xi_o$ for $d_N \rightarrow 0$ (section \ref{sshortdn}).
The decomposition (\ref{Im}) is valid for arbitrary $d_N$ with
$\ep^{j}(y)$ given by Eq. (\ref{epy1}) and can be used to
calculate the current-flux relationship in any NS loop.  At finite
temperature, we find a crossover from paramagnetic to diamagnetic
behavior at small flux when $d_S$ is increased or $T$ lowered, see
section \ref{stempe}.

\section{Long normal segment}\label{slongdn}

\begin{figure}[ht!]
\begin{center}
\epsfxsize 9.0cm \epsffile{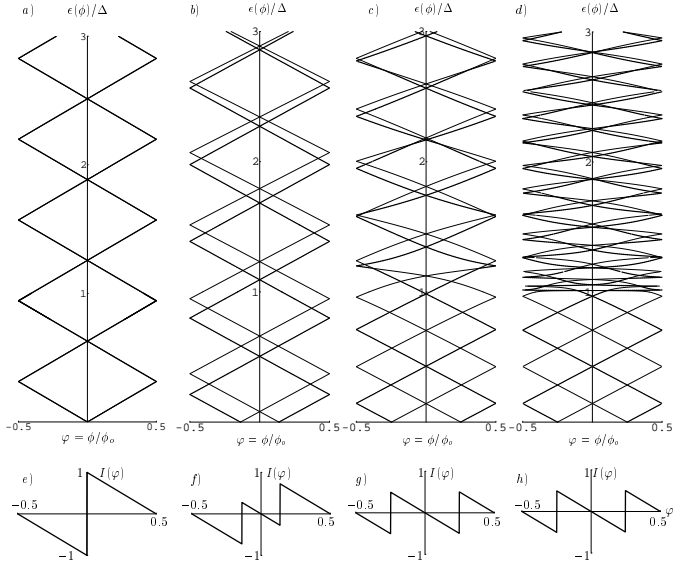}
\end{center}
\caption{NS loop spectrum $\ep^{j}_{\pm}(n,\varphi)$ for a) $d_{S}=0$,
b) $d_{S}=\xi_o$, c) $d_{S}=5\xi_o$ and d) $d_{S}=20 \xi_o$,
keeping $d_{N}$ equal to $10 \xi_{o}$. The number of electrons per
spin direction $N$ is even. Levels with $j=+1$ correspond to the
thin lines and those with $j=-1$ to the thick lines. The currents
e), f), g), h) are plotted in units of $I_{o}=2ev_{F}/d^*$ below
the corresponding spectrum a), b), c), d). }\label{spectredsvar}
\end{figure}

In this section, we study the evolution of the spectrum and of
$I(\phi)$ as a function of $d_S$ for a NS ring with $d_N \gtrsim 2
\xi_o$. We have shown in paragraph \ref{snum} that in this case
the second term in Eq. (\ref{Im}) is negligible.

\subsection{Excitation spectrum  for arbitrary $d_S$}

In Fig. \ref{spectredsvar}(a,b,c,d), we have plotted the
excitation spectrum for $d_{S}=0,\xi_o,5 \xi_o,20\xi_o$ keeping
$d_{N}$ equal to $10 \xi_o$. For the low energy part of these
spectra, the flux dependence is linear with the slope $\pm h
v_{F}/d_N$. For the normal ring $d_{S}=0$, the branches $j=\pm 1$
coincide and we recover the spectrum of the normal ring with the
double degeneracy due to the spin . For finite $d_S$, these
branches are shifted by $\Delta \varphi = \pm \arccos(\cos(k_F
L)/\cosh\lambda d_S )/2\pi$. For $d_S=5 \xi_o$ and $d_S=20 \xi_o$,
they are shifted by half a quantum of flux $\phi_o/2$ . Indeed,
the low energy excitation spectrum is the same as the spectrum of
a normal ring of perimeter $d^*$ and with $\cos k_{F}L$ replaced
by $\cos k_{F}L/\cosh \lambda d_{S}$. The low energy spectra for
$d_{S}=5 \xi_o,20 \xi_o,\infty$ are quite similar and correspond
to quasiparticle motion confined in the normal region of the loop.
The spectrum above the gap is more sensitive to the value of $d_S$
because the level spacing scales as $1/L$ corresponding to a
quasiparticle motion confined in the whole loop of length
$L=d_{S}+d_{N}$.

\subsection{Current-flux relationship for arbitrary $d_S$}

The Fourier coefficients in the harmonics expansion of the current
Eq. (\ref{IFourier}) are $I_m = I^{+}_{m}+I^{-}_{m}$ with:
\begin{equation}
I^{j}_{m}={2 \over {\pi m}} \frac{1}{\phi_o} {d\ep^{j} \over dy}
(y^{j}_{o})  \cos 2 \pi m y^{j}_{o} \label{Imlong}
\end{equation}
and they depend only on the zero energy Andreev level. The
equation (\ref{epy1}) yields:

\begin{equation}
2 \pi y^{j}_{o} = - j \arccos \left({\cos k_ {F}L \over \cosh
\lambda d_S} \right) \label{ytop}
\end{equation}
and $d\ep^{j}/dy= h v_F/d^*=E_A$. The energy $E_A$ is the typical
displacement of one energy level when the flux is varied. This is
also the energy spacing between Andreev levels. The order of
magnitude of the critical current is then $E_{A}/\phi_o$.
According to Eq. (\ref{Im}), the Fourier expansion of the current
for arbitrary $d_S$ reads:
\begin{equation}
I(\phi)={4 \over \pi} {ev_{F} \over d^*} \sum_{m=1}^{\infty}
{T_m(X) \over m} \sin 2 \pi m \phi / \phi_{o} \label{ILnsloop}
\end{equation}
with
\begin{equation}
 X=\frac{\cos k_{F}L}{\cosh \lambda d_S}\label{tm1}
\end{equation}
and:
\begin{equation}
T_m(X)=\cos \left( m \arccos X \right)\label{tm2}
\end{equation}
The coefficients $T_m(X)$ are the m-th order Tchebytchev
polynomials. The parameter X depends both on the band filling and
on the crossover parameter $\lambda d_S$. The first Tchebychev
polynomials are $T_1(X)=X$, $T_2(X)=2 X^{2}-1$,
$T_3(X)=4X^{3}-3X$, .... For a fixed number of electrons per spin
direction, namely for $\cos k_{F}L=\pm 1$, the beginning of the
current expansion is given by the following expression
\cite{footbk}:

\begin{eqnarray}
I(\phi) &=&   {4 \over \pi} {ev_{F} \over d^*} \left[ \pm { 1
\over \cosh \lambda d_{S}} \sin 2 \pi m \phi / \phi_{o} \right.
\nonumber
\\ &+& \left.
{{1-\sinh^{2} \lambda d_{S}} \over {2 \cosh^{2} \lambda d_{S}}}
\sin 4 \pi m \phi / \phi_{o} + ... \right] \label{ILnsloop2}
\end{eqnarray}
The formula above describes the suppression of the first harmonic
$m=1$ when $d_S$ goes to infinity. The suppression of odd
harmonics is a general feature of the crossover from persistent
current in normal loops to Josephson current in SNS junctions.

\subsection{Case $d_{S}\gg\xi_o$: Josephson limit}\label{sslongdnandds}
We recall that for large $d_S$, the NS loop threaded by a magnetic
flux $\phi$ behaves like a SNS junction with a superconducting
phase shift $4 \pi \phi / \phi_o$ between the leads. In this
$d_{S}\rightarrow \infty$ limit, the functions $\ep^{j}(y)$ are
flat outside the gap. Due to the infinite size of the system, the
spectrum becomes a true continuum above the gap. Far from the gap,
the density of levels is given by:
\begin{equation}
\frac{dy}{d\ep} = \frac{d_N}{h v_F} + \frac{d_S}{h v_F}
\frac{\ep}{(\ep^2-\Delta^2)^{1/2}} \label{Dol}
\end{equation}
It is obtained from expressions (\ref{rep2}), (\ref{Thetaup}) and
(\ref{epy1}). The first term is the density of levels in a normal
ring of perimeter $d_N$ at the Fermi level and the second term is
the BCS singularity at $\Delta$. At $\ep \gg \Delta$, the total
density of levels tends to those of a normal loop of perimeter
$L=d_{S}+d_{N}$. Inside the gap for $\ep<\Delta$, we have:
\begin{equation}
2 \pi y^{j}(\ep)={{\ep d_N} \over {\hbar v_F}} - \arccos \frac{
\ep}{\Delta} + (1-j) \frac{\pi}{2}
\end{equation}
which corresponds to the Kulik spectrum (\ref{kuliknj}). It is
dominated by the linear behaviour of the first term for long
junctions $d_{N}\gg\xi_o$, except very close to $\Delta$. Below
the gap, the flux dependent spectrum is similar to those plotted
in Figs. \ref{spectredsvar}(c,d). In this limit, $X \rightarrow 0$
and only even $m=2p$ harmonics are non zero in Eq.
(\ref{ILnsloop}) because $T_{2p+1}(0)=0$. It leads to the
following $\phi_{o}/2$ periodic current:
\begin{equation}
I(\phi)={ 2 \over \pi} {{ev_{F}} \over
{d_{N}+\xi_o}}\sum_{p=1}^\infty {{(-1)^p} \over {p}} \sin 4 \pi p
\phi / \phi_o  \label{ISNS2}
\end{equation}
This is the Fourier expansion of a saw-tooth current-flux
relationship for the purely longitudinal channel of long SNS
junctions. In section \ref{mulcanal}, we show that it corresponds
to the result of Bardeen and Johnson $\cite{Bardeen72}$.

\subsection{Case $d_S=0$: persistent currents in normal rings}

If we remove the superconductor, the parameter $X$ is equal to
$\cos k_F L$ and the effective length $d^*$ is the perimeter $L$.
Then, the model describes the persistent current in a purely
normal ring of length L with fixed chemical potential. We recover
the well-known result \cite{Cheung88}:
\begin{equation}
I(\phi)={4 \over \pi} {ev_{F} \over L}\sum_{m=1}^\infty {{\cos m
k_{F}L} \over m} \sin 2 \pi m \phi /\phi_o  \label{ILnormal}
\end{equation}
the current $I(\phi)$ includes both spin directions. This result
is correct in the framework of the quasi-classical approximation
$\ep \ll E_F$. In fact, Eq. (\ref{ILnormal}) is the zero order
contribution in the $1/k_{F}L$ expansion. The following term in
this expansion is due to the quadratic dispersion relation of free
electrons and is evaluated in appendix \ref{appkfl}.

\section{Long superconducting segment}\label{sshortdn}

We consider now the case of large $d_S$ and we study the crossover
from long to short SNS junctions obtained by decreasing $d_N$. For
vanishing $d_N$, the levels acquire a nonlinear behavior over a
large energy range of order $\Delta$. Then, the two terms in the
decomposition Eq. (\ref{Im}) become of the same order of
magnitude. For $d_{N}=0$, we recover analytically the current
through a short SNS junction.

\subsection{Excitation spectrum for arbitrary $d_N$}

\begin{figure}[ht!]
\begin{center}
 \epsfxsize 9cm
\epsffile{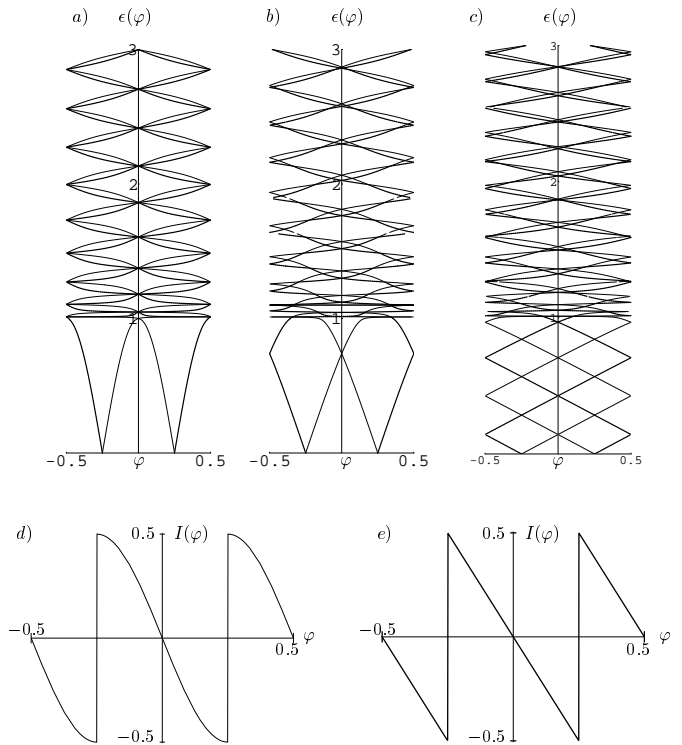}
\end{center}
\caption{NS loop spectrum $\ep^{j}_{\pm}(n,\varphi)$ for $d_{S}=20
\xi_o$ and a) $d_{N}=0$,  b) $d_N=\xi_o$ and c) $d_N=10\xi_o$. The
currents d) and e) are expressed in units of $I_{o}=2 ev_F/d^{*}$
and correspond respectively to the spectra a) and c). Levels with
$j=+1$ correspond to the thin lines and those with $j=-1$ to the
thick lines.} \label{rpvardn}
\end{figure}
On Fig. \ref{rpvardn}(a,b,c), we have plotted the excitation
spectrum for $d_{N}=0,\xi_o,10\xi_o$ and for $d_{S}=20 \xi_o$. For
$d_{N}=0$, there is only one spin degenerate Andreev level in the
gap. As $d_N$ is increased, new Andreev levels appear, see Fig.
\ref{rpvardn}(a) to Fig. \ref{rpvardn}(b) and  Fig.
\ref{rpvardn}(c). As discussed in section \ref{ssexcit2}, the
number of Andreev levels is roughly $4(d_N/\xi_o+1)/\pi$. The
spectrum is linear at low energy and only the last Andreev level
at the vicinity of $\Delta$ varies nonlinearly with the magnetic
flux.

\subsection{Current-flux relationship for $d_N=0$ : short junctions}

In the short junction limit $d_{N}=0$, the eigenvalue equation is:
\begin{equation}
2 \pi y^{j}(\ep)= - \arccos \frac{ \ep}{\Delta} + (1-j)
\frac{\pi}{2}  \label{yshort}
\end{equation}
The function $\ep^{j}(y)$ is plotted in Fig. \ref{epshort}. There
is only one Andreev level in the gap which corresponds
alternalively to $j=1$ or $j=-1$. The spectrum can be written as:
\begin{equation}
\ep(\phi)=\Delta \mid \cos 2 \pi \phi/\phi_o \mid
\label{Andsingle}
\end{equation}
At $T=0$, the current corresponding to this unique Andreev level
is $\phi_{o}/2$ periodic and given by:
\begin{equation}
I(\phi)= - 2 \pi {\Delta \over \phi_o} \sin 2 \pi \phi / \phi_o
\label{IBeena}
\end{equation}
for $|\phi/\phi_{o}|<1/4$, see Fig. \ref{rpvardn}(d). This result
was obtained by Kulik-Omel'yanchuk\cite{KO} and Beenakker and H.
van Houten \cite{Beenakker1}. In appendix \ref{appshortsns}, we
show how to recover this result from our formalism and Eq.
(\ref{Im}). In this case, both terms in this equation contribute
with the same order of magnitude, so that the BJ approximation
clearly breaks down for $d_N=0$.

Finally, we can compare the short $d_N$ case, $\Delta<E_A=h
v_F/d^*$, and the long $d_N$ case, $\Delta>E_A$, developed in
\ref{slongdn}. In both cases, the current is
$\phi_{o}/2$-periodic, diamagnetic at small flux and there is a
jump at $\phi_{o}/4$. In the former case, $I(\phi)$ is triangular
and the critical current is of order $E_A/\phi_o$. In the latter
case, $I(\phi)$ has the sine dependence and the critical current
is of order $\Delta/\phi_o$. It is a particular case of the
well-known statement that the critical Josephson current in a SNS
junction is given by the minimum of the two energy scales $E_A$
and $\Delta$\cite{Likha79}, $E_A$ being the Thouless energy of the
clean NS loop.

\section{Which levels carry the current?}\label{sinout}

\begin{figure}[ht!]
\begin{center}
 \epsfxsize 8.5cm
\epsffile{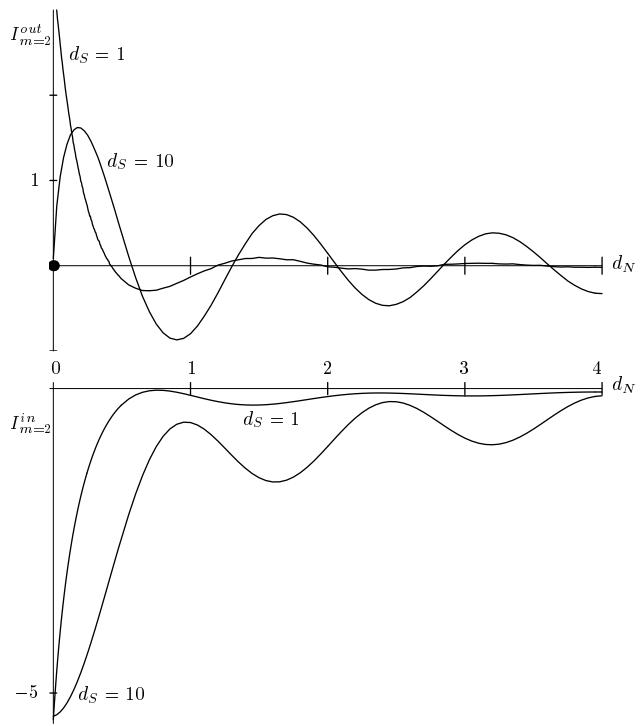}
\end{center}
\caption{Contributions to the second harmonic from states outside
the gap $I_{m=2}^{out}$ and from states below the gap
$I_{m=2}^{in}$ in units of $\Delta/\phi_o$ for $d_{S}=\xi_o$ and
$d_{S}=10 \xi_o$ as a function of $d_{N}$. The dot represents the
expected cancellation of the current carried by states above the
gap in the limit of large $d_S$.} \label{harmo2inout}
\end{figure}

The equilibrium current in a NS loop is carried both by levels
below and above the superconducting gap
$\Delta$\cite{Bagwell}$^,$\cite{Samuel}. Although our formalism is
based on a decomposition between linear vs. nonlinear flux
dependence of the levels, we can separate in each term of Eq.
(\ref{Im}) into contributions from states above and below the gap.
The result is shown on Fig. \ref{harmo2inout}. Although the total
current decreases monotonously with $d_N$, both contributions are
oscillating functions of $d_{N}$. These oscillations in the subgap
current harmonics correspond to the apparition of new Andreev
levels below the gap when $d_N$ increases. Since the number of
Andreev levels is $4 d^*/\pi \xi_0$, the oscillations have
periodicity $\pi \xi_0 /2$. Although we have not checked this
numerically, we believe that the contribution carried by the
states outside the gap cancels whenever the level crossing the gap
has zero slope. Otherwise, when an Andreev level crosses the gap,
it still carries current on the other side of the
gap\cite{Bagwell}$^,$\cite{Samuel}.

For the case of the short SNS junction, $d_N=0$ and $d_S \gg
\xi_o$, the current is solely carried by the single Andreev level.
As seen on Fig. \ref{harmo2inout}, when $d_S$ is close to $\xi_o$,
there is a contribution from states above the gap, which is
significant when $d_S$ is close to $\xi_o$.

\section{Temperature effect}\label{stempe}
We treat the effect of a finite temperature on the crossover from
SNS junctions to normal rings. We essentially focus on the case
$d_N \gtrsim 2 \xi_0$.
\subsection{Harmonics expansion at finite temperature}
We use Eq. (\ref{ImNmT}) to compute the harmonics of the current
at finite temperature. Compared to Eq. (\ref{ILnsloop}), these
harmonics are reduced by the following thermal factor:

\begin{equation}
I_{m}(T)=I_{m}(T=0) {x \over {\sinh x}}
\end{equation}
with $x=2 \pi^{2} m T / E_{A}$ . The characteristic energy scale
associated with the $m$-th harmonic is given by $E_{A}/m$. The
resulting current for $d_N \gtrsim 2 \xi_o$ is:

\begin{equation}
I(\phi,T)=8 \pi \frac{T}{\phi_o} \sum_{m=1}^{\infty}
\frac{T_m(X)}{\sinh \pi m \frac{T}{\Delta}\frac{d^*}{\xi_o}} \sin
2 \pi m \phi / \phi_{o} \label{ILnslooptempe}
\end{equation}
where $X$ and $T_m(X)$ are given by Eqs. (\ref{tm1},\ref{tm2}).

\subsection{Transition from dia- to paramagnetic loops}
\begin{figure}[ht!]
\centerline{ \epsfxsize 9cm \epsffile{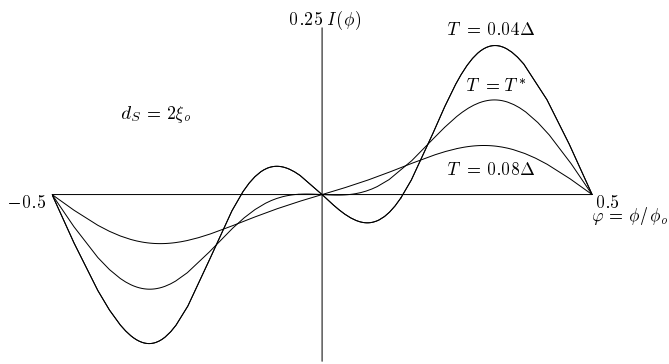}}
\caption{Current-flux relationships for $d_S=2 \xi_o$, $d_N=10
\xi_o $ and for $\cos k_{F}L=1$. At $T=T^*=0.056 \Delta$, the
slope $\partial I/\partial \phi$  at the origin $\phi=0$ changes
in sign. The current is represented in units of
$I_{o}=\Delta/\phi_o$.} \label{tempe1}
 \end{figure}

For $N$ even, a normal ring has a paramagnetic magnetization at
zero flux while a SNS junction carries a diamagnetic current. At
$T=0$, the transition from diamagnetic to paramagnetic behavior
occurs at $d_S=0$. At finite temperature, this transition occurs
at a finite $d_S$. At fixed $d_S$ and $d_N$, there is a similar
crossover as a function of the temperature. In Fig. \ref{tempe1},
we see that the slope at the origin of the $I(\phi)$ curve changes
sign at $T=T^{*}(d_S,d_N)$. The small flux current is diamagnetic
for $T < T^*$ and paramagnetic for $T>T^*$. The function
$T^{*}(d_S,d_N)$ is an increasing function of $d_S$ and a
decreasing function of $d_N$ given by:
\begin{equation}
T^*=\frac{\Delta d_S}{\pi d^*}\label{Ttransi}
\end{equation}
This behavior can be understood from the harmonics expansion Eq.
(\ref{ILnslooptempe}). For $N$ even, the first harmonic is
paramagnetic while the second is diamagnetic. When the temperature
is increased, the reduction of the second harmonic is stronger
than for the first one and the resulting current becomes
paramagnetic. The $I(\phi)$ curve for $N$ odd is obtained by a
$\phi_o/2$ translation of the $I(\phi)$ shown on Fig.
\ref{tempe1}. One sees that the magnetization at small flux is
always diamagnetic. Indeed, all the harmonics are negative for $N$
odd.

\subsection{Ensemble average}\label{saverage}
Here, we calculate the average current of a large assembly of
isolated single-channel NS loops with $d_N \gg \xi_o $ and a fixed
number of particles. Statistically, half of them have an even
number of electrons per spin direction $N$ and for the others $N$
is odd. As we have seen in the previous paragraph, a transition
from diamagnetism to paramagnetism occurs for the rings with even
$N$, while the rings with $N$ odd are always diamagnetic. Does the
total orbital magnetism of the assembly exhibits a transition?  We
know that an assembly of normal rings has paramagnetic
magnetization at small flux and is
$\phi_{o}/2$-periodic\cite{mtb}. In the opposite limit
$d_{S}/\xi_{o}\gg 1$, the current in each loop is diamagnetic and
$\phi_{o}/2$ periodic. The transition occurs at a very small value
of $d_S$ close to $0.5 \xi_o$. In conclusion, we find that the
current is always diamagnetic for values of $d_S$ and $d_N$ above
$\xi_o$.

\section{Multichannel rings}\label{mulcanal}
Up to now, we have considered a single-channel NS ring. From now
on, we extend our study to multichannel NS rings. First, we
present spectra with a small number of transverse channels and
follow the evolution of the different channels when $d_N$ and
$d_S$ are varied. Secondly, we study the crossover from Josephson
current to persistent current in the clean multichannel NS loop at
finite temperature.

\begin{figure}[ht!]
\centerline{ \epsfxsize 9cm \epsffile{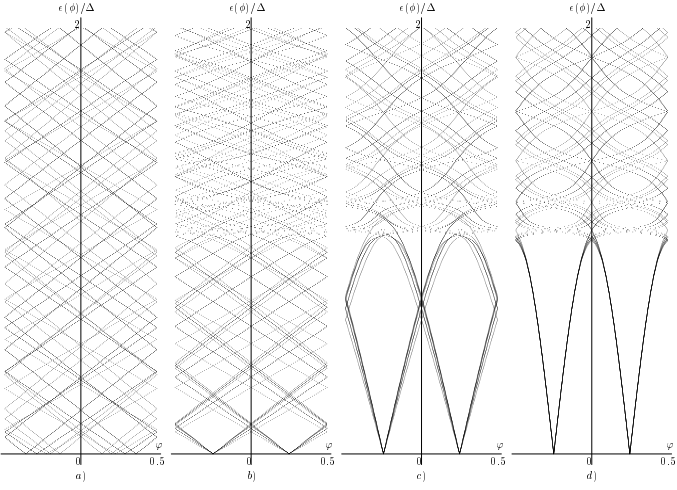}}
\caption{Multichannel spectra. The cross section is a square of
size $(2 \lambda_F)^2$. For clarity, we have represented only a
few flux-dependent channels among the $M=k_{F}^2 S/4 \pi$
transverse channels. a) $d_S=\xi_o$ and $d_N=10 \xi_o$, b)
$d_S=10\xi_o$ and $d_N=10 \xi_o$, c) $d_S=10 \xi_o$ and
$d_N=\xi_o$, d) $d_S=10 \xi_o$ and $d_N=0$.} \label{multi}
 \end{figure}

The spectrum of clean multichannel rings can be obtained
straightforwardly since the different channels are decoupled and
characterized by their momenta ${k_{y},k_{z}}$ quantized along
transverse directions $y$ and $z$. The spectrum of each of these
channels is simply obtained by the substitutions $v_{F}\rightarrow
v_{Fx}$ and $k_{F}\rightarrow k_{Fx}$ in the Eqs.
(\ref{Bktrig},\ref{epy1}). As an example, we have represented on
Fig. \ref{multi} the spectra of a ring with a square section of
size $(2 \lambda_F)^2$ for different values of $d_N$ and $d_S$. As
$d_N \rightarrow 0$, the spectra of the different channels all
shrink towards the single-channel spectrum and become completely
degenerate when $d_N=0$. As a consequence, the $I(\phi)$ for a
$M$-channel short junctions $d_N=0$ is exactly $M$ times the
single-channel result (\ref{IBeena}).
In long SNS junctions, the current $I(\phi)$ carried by any
transverse channel has the same flux dependence as the
longitudinal one, namely a triangular $I(\phi)$ with current jumps
at $\phi=\pm \phi_o/4$. Nevertheless, the critical current is
different for each channel as we can see from the spectrum Fig.
\ref{multi}, because the slopes $-\partial \ep/\partial
\phi(\ep=0)$ are different. Finally, the total current stays
proportional to the number of transverse channels $M$.

 The total current is the sum of the single-channel currents. For
$d_N \gtrsim 2\xi_o$, it is given by:
\begin{equation}
I(\phi)={ 4 \over \pi} \sum_{k_{y},k_{z}} {{ev_{Fx}} \over
{d_{N}+\xi_o}}\sum_{m=1}^\infty {{T_{m}(X)} \over {m}} \sin 2 \pi
m \varphi  \label{multi1}
\end{equation}
where:
\begin{equation}
 X=\frac{\cos k_{Fx}L}{\cosh \lambda d_S} \label{Xth}
\end{equation}
In the case of a NS ring with many transverse channels, the
discrete sum Eq. (\ref{multi1}) over transverse channels can be
replaced by an integral. Each channel carries a current given by
Eq. (\ref{ILnslooptempe}) with $v_{F} \rightarrow v_{Fx}=v_{F}
\cos \theta$, where $\theta$ is the angle between the direction
$x$ and the Fermi momentum of the channel. By counting of the
$(k_{y},k_{z})$ satisfying $k_{x}^2+k_{y}^2+k_{z}^2=k_{F}^2$ for a
given incidence $\theta$, we obtain the total current:

\begin{equation}
I(\phi)=16 \pi M \frac{T}{\phi_o}\int_{o}^{\pi/2} d\theta
\cos^{2}\theta \sum_{m=1}^\infty {{T_{m}(X)} \over {\sinh\pi m
\frac{T}{\Delta} \frac{d^*}{\xi_o}}} \sin 2 \pi m \varphi
 \label{multi2}
\end{equation}
where $M=k_{F}^2 S/4 \pi$ is the number of transverse channels. In
the limit $d_S \gg \xi_o$ and at $T=0$, Eq. (\ref{multi2}) leads
to:
\begin{equation}
I(\phi)=\frac{4 M}{3 \pi} \frac{ev_{F}} {d_{N}+\xi_o}
\sum_{p=1}^\infty {{(-1)^p} \over {p}} \sin 4 \pi p \varphi
 \label{multi3}
\end{equation}
The corresponding critical current is proportional to $M$:
\begin{equation}
I_{m}=\frac{2 M}{3} \frac{ev_{F}} {d_{N}+\xi_o} \label{multi4}
\end{equation}
This is the old result found by Bardeen and
Johnson\cite{Bardeen72}. Ishii\cite{Ishii70} and Svidsinski {\it
et al.}\cite{Svid72} found different numerical pre-factors.

\begin{figure}[ht!]
\centerline{ \epsfxsize 9cm \epsffile{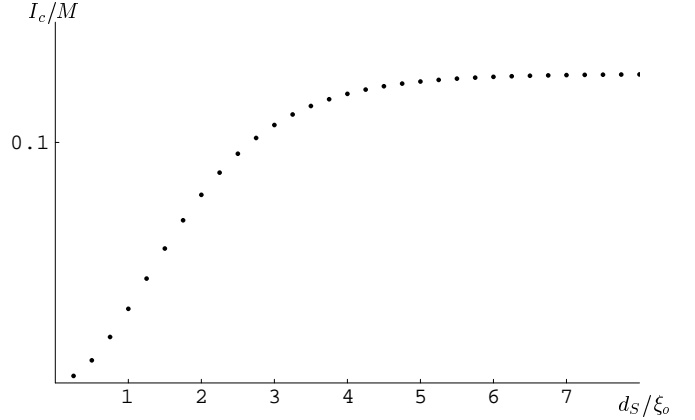}} \caption{Critical
current per transverse channel as a function of $d_S$ for $d_N=10$
and $k_F \xi_o \gg 1$ at $T=0.03 \Delta$.} \label{multicri}
 \end{figure}

In the opposite limit of the multichannel normal ring $d_S=0$, the
total current averages to zero \cite{multiibm} for large $k_F
\xi_o$. In Fig. \ref{multicri}, we show the the crossover of the
critical current from $M e v_F /(d_N+\xi_o)$ for large $d_S$ to
zero as $d_S$ approaches zero. This can be understood from the
spectra Fig. \ref{multi}. Indeed, when $d_S$ is large, the low
energy spectra of each channel are in phase with $\ep=0$ for
$\phi=\pm \phi_o/4$. When $d_S \rightarrow 0$, there is finite
dephasing between the different spectra given by $\Delta \varphi=
1/2\pi . \arccos X$ which leads to a cancellation of the total
current.

\section{Conclusion}\label{conclu}
We have calculated the full excitation spectrum of a NS loop
threaded by an Aharonov-Bohm flux, for any value of $d_N$ and
$d_{S}$: in particular, we show for the first time the spectrum of
the NS loop above the gap. We have identified the contribution to
the current directly originating from the nonlinearities in the
flux dependent spectrum. We recover known results for short and
long SNS junctions when $d_{S}\gg \xi_o$. For the single-channel
NS ring at zero temperature, we recover the result of BK and our
method allows us to clarify and justify the approximation made by
BK. Moreover, we extend the study of the NS loop to the finite
temperature and to the multichannel cases.

We would like to thank M. B\"uttiker, H. Bouchiat and S. Gueron
for useful and stimulating discussions.

\section{Appendixes}

\subsection{NS loop excitation spectrum} \label{app1}
We consider a purely one dimensionnal NS loop with a
superconducting segment in the region $0<x<d_S$ and a normal
segment in $d_{S}<x<L$. This loop is threaded by a magnetic
Aharonov-Bohm flux $\phi$.

We first investigate states with energies below $\Delta$. In the
normal metal, the quasiparticles wavefunctions are plane waves
oscillating at wavevectors $k_{e/h}=k_{F} \pm \ep/\hbar v_{F}$
just below (hole solution) or just above (electron solution) the
Fermi momentum $+k_F$:
\begin{equation}
\left(
\begin{array}{c}
 u (x)\\ v(x)
\end{array}
\right) = A \left(
\begin{array}{c}
1\\ 0
\end{array}
\right) e^{ik_{e}x} + B
 \left(
\begin{array}{c}
0 \\  1
\end{array}
\right)
 e^{ik_{h}x} \label{normal}
\end{equation}
In the superconductor , they are:
\begin{equation}
\left(
\begin{array}{c}
 u (x)\\ v(x)
\end{array}
\right) = C
 \left(
\begin{array}{c}
e^{i \eta_{\ep}}\\e^{-i \eta_{\ep}}
\end{array}
\right) e^{(ik_{F}-\lambda_{\ep})x} + D
 \left(
\begin{array}{c}
e^{-i \eta_{\ep}} \\  e^{i \eta_{\ep}}
\end{array}
\right) e^{(ik_{F}+\lambda_{\ep})x} \label{supra}
\end{equation}
where $e^{2i \eta_{\ep}}=( \ep + i\sqrt{\Delta^2-\ep^2})/\Delta$
and $\lambda_{\ep}=\sqrt{\Delta^2-\ep^2}/\hbar v_F$. We express
the continuity of these functions at the NS interfaces. In
presence of a reduced flux $\varphi=\phi/\phi_o$ it leads to the
system:
\begin{eqnarray}
A e^{ik_{e} L} &=& e^{2 \pi i \varphi} ( e^{i \eta_{\ep}}
C+e^{-i\eta_{\ep}} D) \nonumber \\B e^{ik_{h} L} &=& e^{- 2 \pi i
\varphi}(e^{-i \eta_{\ep}} C+ e^{i \eta_{\ep}} D) \nonumber \\ A
e^{ik_{e}d_{S}}&=& e^{(ik_{F}-\lambda_{\ep})d_{S}} e^{i
\eta_{\ep}} C + e^{(ik_{F}+\lambda_{\ep})d_{S}} e^{-i \eta_{\ep}}D
\nonumber \\ B e^{ik_{h} d_{S}}&=& e^{(ik_{F}-\lambda)d_{S}} e^{-i
\eta_{\ep}} C + e^{(ik_{F}+ \lambda)d_{S}} e^{i \eta_{\ep}} D
\label{systNS}
\end{eqnarray}
The continuity of the derivatives is automatically satisfied in
the quasi-classical approximation $\ep \ll E_F$. In this
approximation, there is no mixing between excitations around
$+k_F$ and excitations around $-k_{F}$. For this reason, it was
possible to consider only solution oscillating around $+k_F$ in
the ansatz Eqs. (\ref{normal},\ref{supra}). The determinant of the
system Eq. (\ref{systNS}) gives the eigenvalues equation:
\begin{equation}
2 i \sin 2 \eta_{\ep} \cos k_{F}L = e^{2 \pi i\varphi - i
\frac{\ep d_{N}}{\hbar v_{F}}} \sinh ( \lambda d_{S} + 2 i
\eta_{\ep} ) - c.c. \label{g2}
\end{equation}
which is identical to the spectrum obtained by BK:
\begin{eqnarray}
\cos k_{F}L \sin 2\eta_{\ep}&=&\sin 2\eta_{\ep} \cosh
\lambda_{\ep} d_S \cos \left({{\ep d_N} \over {\hbar v_F}} - 2 \pi
\varphi \right) \nonumber\\ && -\cos 2\eta_{\ep} \sinh
\lambda_{\ep} d_S \sin \left({{\ep d_N} \over {\hbar v_F}} - 2 \pi
\varphi \right) \nonumber
\end{eqnarray}
This formula can be reduced in the following form:
\begin{equation}
\cos k_{F}L=r_{\ep} \cos \left(\frac{\ep d_N}{\hbar v_F}- 2 \pi
\varphi + \Theta_{\ep} \right)\label{Bkapp}
\end{equation}
with the parameters $r_{\ep}$ and $\Theta_{\ep}$ defined by:
\begin{eqnarray}
r_{\ep} e^{i \Theta_{\ep}}&=&\frac{\sinh(2 i \eta_{\ep}-
\lambda_{\ep} d_{S})}{\sinh 2 i \eta_{\ep}} \\ &=& \cosh
\lambda_{\ep} d_{S} + i \cot 2 \eta_{\ep} \sinh \lambda_{\ep}
d_{S}
\end{eqnarray}
One expression for the modulus $r_{\ep}$ is:
\begin{equation}
r_{\ep}=\mid \cosh \lambda_{\ep} d_{S}+ i \cot 2 \eta_{\ep} \sinh
\lambda_{\ep} d_{S} \mid \label{r1}
\end{equation}
and the phase $\Theta_\ep$ is satisfies:
\begin{equation}
\tan \Theta_{\ep}= \cot 2 \eta_{\ep} \tanh \lambda_{\ep} d_{S}
\label{t1}
\end{equation}

We now look for quasiparticle states with energies above $\Delta$.
In the normal region, the form of the wavefunctions is unchanged.
In the superconductor, they become:
\begin{equation}
\left(
\begin{array}{c}
 u (x)\\ v(x)
\end{array}
\right) = C
 \left(
\begin{array}{c}
e^{-\delta_{\ep}} \\ e^{\delta_{\ep}}
\end{array}
\right) e^{i(k_{F}-\delta k_{\ep})x} + D
 \left(
\begin{array}{c}
e^{\delta_{\ep}} \\e^{-\delta_{\ep}}
\end{array}
\right)
 e^{i(k_{F}+\delta k_{\ep})x} \label{supra2}
\end{equation}
where $e^{2 \delta_{\ep}}=(\ep + \sqrt{\ep^2-\Delta^2} )/\Delta$.
Therefore, the eigenvalue equation can be obtained directly from
the preceding study with the replacement $-i \eta_{\ep}
\rightarrow \delta_\ep$ and $\lambda \rightarrow i \delta k_\ep$.
\begin{figure}[ht!]
  \centerline{\epsfxsize 6cm
  \epsffile{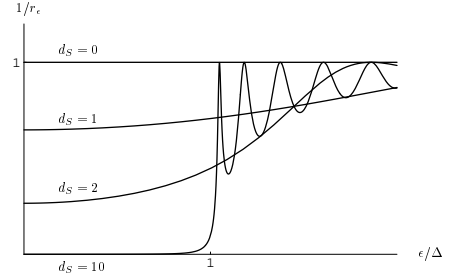}}
  \caption{$1/r_\ep$ as a function of energy for different values of $d_S$ in units of $\xi_o$.}\label{repsi}
\end{figure}
\begin{figure}[ht!]
  \centerline{\epsfxsize 6cm
  \epsffile{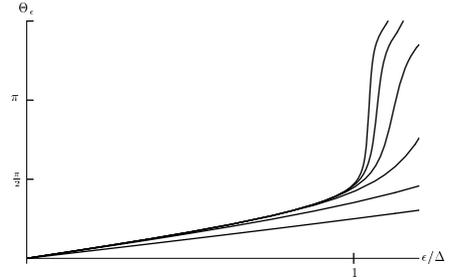}}
  \caption{Phase shift $\Theta_{\ep}$ as a function of energy for different values of $d_{S}$. The curves correspond to $d_{S}=1,2,4,6,8,10$ in units of $\xi_o$
from bottom to top. For $d_{S} \rightarrow \infty$, $\Theta_{\ep}
\rightarrow \pi/2-\arccos(\ep/\Delta)$.}
  \label{figthet1}
\end{figure}
\begin{figure}[ht!]
  \centerline{\epsfxsize 6cm
  \epsffile{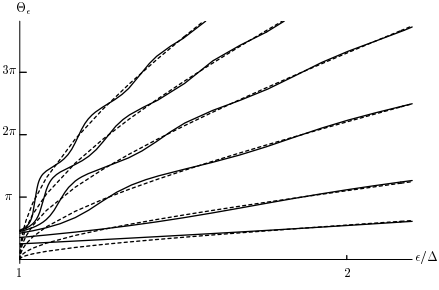}}
  \caption{Asymptotic behaviour of $\Theta_\ep$ above $\Delta$ for different values
of $d_S$. The curves correspond to $d_{S}=1,2,4,6,8,10$ in units
of $\xi_o$ from bottom to top.  At large energies, $\Theta_\ep$ is
close to $\delta k_{\ep} d_{S}$ which is represented in dashed
lines for each value of $d_S$.} \label{figth2}
\end{figure}
We obtain the Eq. (\ref{Bkapp}) with the complex parameter:
\begin{eqnarray}
r_{\ep} e^{i \Theta_{\ep}}&=&\frac{\sinh(i \delta k_{\ep} d_{S}+ 2
\delta_{\ep})}{\sinh 2 \delta_{\ep}}\\ &=& \cos \delta k_{\ep}
d_{S} + i \coth 2 \delta \sin \delta k_{\ep} d_{S}
\end{eqnarray}
The modulus is given by:
\begin{equation}
r_{\ep}=\mid \cos \delta k_{\ep} d_{S}+ i \coth 2 \delta \sin
\delta k_{\ep} d_{S} \mid \label{r2}
\end{equation}
and the phase by:
\begin{equation}
\tan \Theta_{\ep}= \coth 2 \delta_{\ep} \tan \delta k_{\ep} d_{S}
\label{t2}
\end{equation}
Equations (\ref{r2},\ref{t2}) can be obtained directly from
(\ref{r1},\ref{t1}) by prolongation to energies $\ep
> \Delta$.
The fonctions $r_\ep$ and $\Theta_\ep$ are plotted in Figs.
(\ref{repsi},\ref{figthet1},\ref{figth2}) for different values of
$d_{S}$. We note that $r_{\Delta} e^{i \Theta_\Delta} =1+i
d_{S}/\xi_{o}$. We have seen in the paragraph \ref{ssexcit3} that
these functions have simple limits for very large and for very
small $d_{S}/\xi_o$.

\subsection{Supercurrent in short SNS junctions}\label{appshortsns}
\begin{figure}[ht!] \centerline{ \epsfxsize 6cm
\epsffile{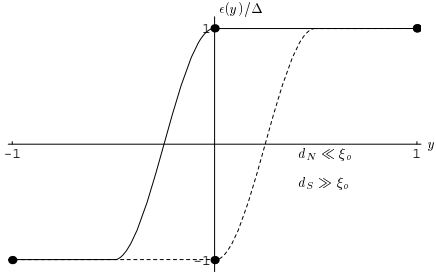}} \caption{$\epsilon^{j}(y)$ for $j=+1$ (solid
line) and $j=-1$ (dashed line) $d_{S}=\infty$, $d_{N}=0$.}
\label{epshort}
\end{figure}

In the case of short SNS junctions, the eigenvalue equation Eq.
(\ref{yshort}) can be inverted. For the $j=1$, we obtain
$\ep=\Delta \cos 2 \pi y $ in the interval $-1/2<y<-1/4$ and for
$j=-1$, we have $\ep=-\Delta \cos 2 \pi y $ in the interval
$0<y<1/4$. These functions $\ep^{j}(y)$ are shown in Fig.
\ref{epshort}.
Now, we detail the calculation of the current for the $j=1$ case.
From Eq. (\ref{Im}), we get:
\begin{equation}
I_{m}^{j=1}=\frac{4 \Delta}{\phi_o} \frac{1}{m} \left[\cos
\frac{\pi m}{2} + 2 \pi \int_{-1/2}^{-1/4} dy \cos 2 \pi m y \cos2
\pi y\right] \label{Ishort2}
\end{equation}
Odd harmonics with $m \geq 3$ are zero. Even harmonics $m=2 p$
are:
\begin{equation}
I_{2 p}^{j=1}=\frac{4 \Delta}{\phi_o} (-1)^p \frac{2 p }{4p^2-1}
\label{Ishort5}
\end{equation}
Similar calculation for $j=-1$, give the same result for $I_m$
when $m\geq2$. The case $m=1$ has to be considered separately and
it is easy to see that the $j=+1$ and $j=-1$ cancel each other.
Consequently, the current is given by:

\begin{equation}
I(\phi)={4 \Delta \over \phi_o} \sum_{p=1}^\infty (-1)^p {p \over
{p^2-1/4}} \sin 4 \pi p \phi / \phi_o  \label{ipshort}
\end{equation}
which is nothing but the harmonics expansion of Eq.
(\ref{IBeena}).

\subsection{Exact calculation of  the persistent currents in a
normal loop}\label{appkfl}

In this appendix, we show how the method developed in section \ref{sdecomp}
 to calculate the harmonics of the current can be used to
obtain the current of the purely normal ring. In this case, it is
easier the equilibrium single particle states spectrum rather than
the excitation spectrum. For a free electron in a ring of length
$L$, the electronic levels are:
\begin{equation}
\ep_{n}(\phi)=\frac{2 \pi^2 \hbar^2}{m_{el} L^2} (n+\varphi)^2
\end{equation}

The derivation of Eq. (\ref{Im}) is still valid for the
equilibrium spectrum.

\begin{eqnarray}
I_{m}(T&=&0)= {4 \over {\pi m}} \frac{1}{\phi_o} \left[ {d\ep
\over dy} (y_{F}) \cos 2 \pi m y_{F}\right. \nonumber \\&-& \left.
\int_{0}^{ y_{F}} dy \frac{d^2\ep}{d^2y} \cos 2 \pi m y \right]
\label{Imeq}
\end{eqnarray}
We have added an additional factor $2$ for the spin degeneracy.
The first term in Eq. (\ref{Imeq}) gives the above result Eq.
(\ref{ILnormal}) but the second term give a non vanishing
correction due to the parabolic dispersion relation. This
correction is easy to evaluate since the curvature $d^2\ep/d^2y=4
\pi^2 \hbar^2 / m_{el} L^2 $ is a constant and we obtain:
\begin{equation}
\delta I (\phi)=-{4 \over \pi} {ev_{F} \over L}
\frac{1}{k_{F}L}\sum_{m=1}^\infty \frac{\sin m k_{F}L}{m^2} \sin 2
\pi m \phi / \phi_o  \label{corrkfl}
\end{equation}
which is the $1/k_{F}L$-corrective term to the $0^{th}$ order term
obtained in Eq. (\ref{ILnormal}).


\begin{thebibliography}{99}
\bibitem{numberel} the results obtained in this paper depend on the parity
of the number $N$ of electrons per spin direction. The total
number of electrons $2N$ is always even throughout this paper.
\bibitem{cohlength} we choose to define the superconductor coherence length by $\xi_{o}=\hbar v_{F}/\Delta$ for
convenience, instead of the usual definition $\xi_{o}=\hbar
v_{F}/\pi \Delta$.
\bibitem{footbk} we use the usual convention for
the current in a loop, namely positive if the current is
paramagnetic. Our result differs by BK one by a factor 4. We think
that our prefactor is right result because it leads to the normal
ring result for $d_S\rightarrow0$. The harmonics expansion follows
directly from the application of our formalism, while in the work
of BK it was constructed by analogy with the long SNS junction
result.
\end{thebibliography}
\end{document}